\documentclass[final]{elsarticle}

\usepackage{hyperref}

\usepackage{color}
\usepackage[normalem]{ulem}
\usepackage{soul}

\usepackage[utf8]{inputenc}

\journal{Journal of \LaTeX\ Templates}

\begin{document}

\begin{frontmatter}

\title{Fractional Quantum Heat Engine}

\author{Ekrem Aydiner} 
\address{Department of Physics, Faculty of Science, \.{I}stanbul University, 34134, \.{I}stanbul, Turkey}

\cortext[mycorrespondingauthor]{Corresponding author}
\ead{ekrem.aydiner@istanbul.edu.tr}




\begin{abstract}
In this work, we introduce the concept of the fractional quantum heat engine. We examine the space-fractional quantum Szilard heat engine as an example to show that the space-fractional quantum heat engines can produce higher efficiency than the conventional quantum heat engines.
\end{abstract}


\end{frontmatter}


\section{Introduction}

The quantum version of the Szilárd engine was proposed by Kim et al.\,\cite{Kim2011} as the counterpart of the classical one \cite{Szilard1929}. In many theoretical studies, it has been shown that the positive work and efficiency can be obtained by using quantum Sizilard engine   \cite{Kim2012,Li2012,Cai2012,Zhuang2014,Bengtsson2018,Park2013}. On the other hand, many experimental realization have been implemented both of quantum engine and quantum Szilard engine  \cite{Koski2014,Serreli2007,Raizen2009,Bannerman2009,Koski2015,Koski2014b,Rosnagel2016,Tercas2017,Jarillo2016,Long2016,Erman2016,Martinez2016,Dinis2016,Agarwal2013}.  Recently, a similar version of the quantum engine has been studied by Thomas et al. \cite{Thomas2019} for the infinite well potential. In this study, authors show that the work and efficiency of this quantum Szilárd engine are positive for single and many particles. 

However, we know quantum heat engines and information theory have not been comprehensively studied by using fractional methods in the literature so far. The first study related to the fractional heat engine was suggested in Ref.\,\cite{Aydiner2021}. In this study, the work extraction and efficiency of the quantum Szilard engine has been discussed for fractional power-law potential. The fractional dynamics play an important role in physics and other science areas \cite{Oldham1974,Miller1993,Shlesinger1993,Klages2008,Sokolov2002,Metzler2000,Hilfer2000,Laskin2000,Laskin2002,Laskin2018}. Therefore, it is important to discuss and understand the role of the fractional dynamics in the quantum heat engines. In the previous study \cite{Aydiner2021}, we obtained the work and efficiency of the model without using fractional calculus tools. However, in the present study, we discuss the fractional quantum Szilard engine as an example and we show that fractional exponent plays an important role on the work and efficiency. Therefore, in this study we consider quantum Szilard engine and we discuss the work and efficiency base on the space-fractional Schrödinger equation for infinite well potential. 

This paper is organized as follows: In Section 2, we summarize the solution of the space-fractional Schrödinger equation for the infinite potential well. In Section 3, we discussed  the theoretical framework of the thermodynamics cycles for the fractional quantum Szilard engine. In Section 4, we obtain work and efficiency of the fractional quantum Szilard engine. We also discussed the effects of the space-fractional exponent on the work and the efficiency. Finally, in the last Section, we summarize the obtained results and give a brief discussion.

\section{Fractional Theory for Infinite Well}

We consider here one-dimensional fractional Schrödinger equation for the infinite well potential. This problem was discussed by Laskin in a several papers \cite{Laskin2000,Laskin2002,Laskin2018}. Here we briefly review the fractional Schrödinger equation and its results for infinite well potential. In the Refs.\,\cite{Laskin2000,Laskin2002,Laskin2018,Wei2015}, the Hamiltonian of the system is given by 
\begin{eqnarray} \label{f-Hamiltonian}
	H_{\alpha} \psi_{n}(x) = E_{\alpha n} \psi_{n},(x) \quad n = 0,1,2,...
\end{eqnarray}
where $\alpha$ denotes the fractional index $0 < \alpha \le 2$. 
For the infinite square well with length of $2a$, the space-fractional Schrödinger equation is given by
\begin{eqnarray} \label{f-Schr}
	D_{\alpha} \left(  - \hbar^{2} \frac{d^{2}}{dx^{2}}  \right)^{\alpha /2}  \psi_{n}(x)  +  V_{i} \psi_{n}(x)  =  E_{\alpha} \psi_{n}(x) 
\end{eqnarray}
where the coefficient $D_\alpha = \chi m c^{2} / (mc)^{\alpha}$ with $\chi$ a positive real number, and $c$ is the speed of the light. When $\alpha=2$, taking $\chi=1/2$ and hence $D=1/(2m)$ \cite{Wei2015}. The first term in Eq.\,(\ref{f-Schr}) corresponds to the fractional kinetic energy 
\begin{eqnarray} \label{f-kinetic}
	T_{\alpha} = D_{\alpha} |p|^{\alpha} =\frac{1}{2} m c^{2} \left(  \frac{ |p| }{m c}  \right)^{\alpha} = D_\alpha  \left(  - \hbar^{2} \frac{d^{2}}{dx^{2}}  \right)^{\alpha /2}  .
\end{eqnarray}
 On the other hand, the potential for infinite well is defined as
\begin{eqnarray} \label{f-potential}
 V_{i} = \left\{ \begin{array}{ll}
	0 & \mbox{if $|x| < a$} \\
	+\infty & \mbox{if $|x| > a$} \end{array} \right.  
\end{eqnarray}
the potential takes infinite value at the $x=\pm a$. Finally, the solutions of the fractional Schrödinger equation are given by 
\begin{eqnarray} \label{f-Wave}
\psi_{n} (x)  = \left\{ \begin{array}{ll}
	\frac{1}{\sqrt{2}}\sin\frac{n\pi}{2a}x + a& \mbox{if $|x| < a$} \\
	0 & \mbox{if $|x| > a$} \end{array} \right.
\end{eqnarray}
and 
\begin{eqnarray} \label{f-energy}
	E_{\alpha n} =  D_{\alpha} \left(    \frac{n \pi \hbar} {2 a}    \right)^{\alpha} 
\end{eqnarray}
where $\psi_{n} (x)$ denotes the wave function solution and $	E_{\alpha n}$ is the energy eigenvalues of Eq.\,(\ref{f-Schr}). Here we note that some authors claim that the solution does not satisfies ground state solution of the space-fractional Schrödinger equation (\ref{f-Schr}). However, it is shown that this problem can be solved by assuming that the solution is the limit of the finite square well problem \cite{Wei2015}. Thus, the solution 
\begin{eqnarray} \label{f-limit}
\lim_{V_{0}\rightarrow \infty} \psi_{n}^{finite} (x)  =  \psi_{n} (x) \qquad \lim_{V_{0}\rightarrow \infty} E_{\alpha n}^{finite}  =  E_{\alpha n}
\end{eqnarray}
satisfies the ground state requirement of the infinite square well \cite{Wei2015}. 

To carry out the work and efficiency of the fractional quantum Szilard engine we need the canonical partition function and heat exchanges for all stages. Therefore, in the next,   we discuss these quantities for the infinite well potential given in Eq.(\ref{f-potential}) following the method given in Ref.\,\cite{Thomas2019}.

\section{Thermodynamics Cycles of the Fractional Engine }

\begin{figure} [ht!]
	\centering	
	\includegraphics[width=10cm,height=6cm]{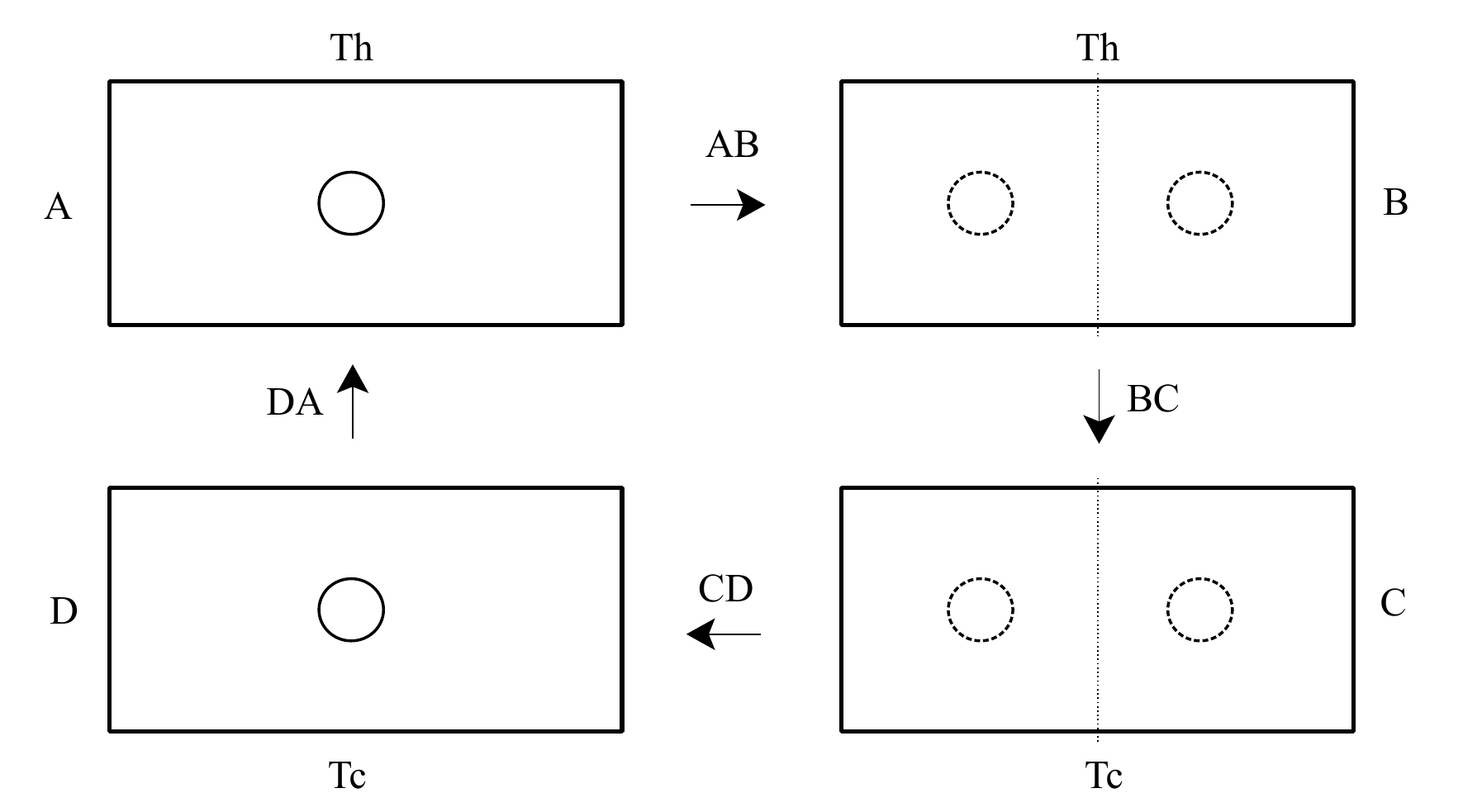}	
	\caption{Work and efficiency of the Szilard quantum heat engine for the infinite well in a box of length $2a$. In this schema boxes A and B interact a heat bath reservoir, boxes C and D interact a cold bath reservoir. The stages AB and CD represent the isothermal insertion and removal processes. The stages BC and DA denote the isochoric processes.}
	\label{f-cycle}
\end{figure}

In the Fig.\,\ref{f-cycle}, box A with a single particle in the infinite potential interacts with a heat bath with a higher temperature of $T_{h}$. However, in AB isothermal stage, a thin potential barrier is completely inserted in the box at the same temperature and the box is divided by the quasi-static insertion process into two half side as in box B which behaves like the two double well potentials. 
In BC isochoric stage, the system is contacted to a heat bath with a lower temperature of $T_{c}$. However, in CD isothermal stage, the inserted thin barrier is again removed. After CD process, the particle is again under a single infinite well potential in box D with a cold bath temperature of $T_{c}$. In the final DA isochoric stage, the system is contacted to a heat bath with a higher temperature of $T_{h}$ and cycle is completed.

The energy eigenvalue for the particle in box A and D is given by
\begin{eqnarray} \label{energy-A}
	E_{\alpha n}^{A,D} =  D_{\alpha} \left( \frac{n \pi \hbar} {2 a}    \right)^{\alpha} .
\end{eqnarray}
The canonical partition functions $Z_{A}(\alpha)$ and $Z_{D}(\alpha)$ for the particle in boxes A and D
\begin{eqnarray} \label{partition-A}
	Z_{A}(\alpha) = \sum_{n=1}^{\infty} \exp \left[-\frac{D_{\alpha} }{k_{B}T_{h}} \left( \frac{n \pi \hbar} {2 a}    \right)^{\alpha} \right] .
\end{eqnarray}
and 
\begin{eqnarray} \label{partition-D}
	Z_{D}(\alpha) = \sum_{n=1}^{\infty} \exp \left[-\frac{D_{\alpha} }{k_{B}T_{c}} \left( \frac{n \pi \hbar} {2 a}    \right)^{\alpha} \right] \ .
\end{eqnarray}
On the other hand, the energy eigenvalue for boxes B and C is given as
\begin{eqnarray} \label{energy-B}
	E^{B}_{\alpha n} = D_{\alpha} \left( \frac{2 n \pi \hbar} {2 a}    \right)^{\alpha} .
\end{eqnarray}
The correspondence partition functions  $Z_{B}(\alpha)$ and $Z_{C}(\alpha)$ are given as
\begin{eqnarray} \label{partition-B}
	Z_{B}(\alpha) = \sum_{n=1}^{\infty} 2 \exp \left[ - \frac{D_{\alpha}}{k_{B}T_{h}} \left( \frac{2 n \pi \hbar} {2 a}    \right)^{\alpha} \right] .
\end{eqnarray}
and
\begin{eqnarray} \label{partition-C}
	Z_{C}(\alpha) = \sum_{n=1}^{\infty} 2 \exp \left[ - \frac{D_{\alpha}}{k_{B}T_{c}} \left( \frac{2 n \pi \hbar} {2 a}    \right)^{\alpha} \right] 
\end{eqnarray}
where the pre-factor 2 in
Eqs.\,(\ref{partition-B}) and (\ref{partition-C}) is written since the boxes are divided into two by the barrier which leads to energy levels two-fold degenerate. 

On the other hand, the heat exchanges $Q_{AB}$, $Q_{BC}$, $Q_{CD}$ and $Q_{DA}$ for all stages of the thermodynamical cycle can be obtained by using the partition functions. For this isothermal process AB, the heat exchanged $Q_{AB}$ can be obtained as
\begin{eqnarray} \label{heat-A-B}
	Q_{AB} = U_{B} - U_{A} + k_{B} T_{h} \ln \frac{Z_{B}(\alpha)}{Z_{A}(\alpha)} 
\end{eqnarray}
where $U_{A}$ and $U_{B}$ are the internal energies of the box A and B, respectively, which can be obtained from
\begin{eqnarray} \label{inertial-A-B}
	U_{A,B} = - \frac{\partial Z_{A,B}(\alpha)}{\partial \beta_{h}}
\end{eqnarray}
where $\beta_{h}=1/k_{B}T_{h}$. Another isothermally stage CD leads to heat exchange $Q_{CD}$ which is written as 
\begin{eqnarray} \label{heat-D-C}
	Q_{CD} = U_{D} - U_{C} + k_{B} T_{c} \ln \frac{Z_{D}(\alpha)}{Z_{C}(\alpha)}
\end{eqnarray} 
where $U_{C}$ and $U_{D}$ are internal energies for box C and
D, respectively which can be found from
\begin{eqnarray} \label{inertial-C}
	U_{C,D} = - \frac{\partial Z_{C,D}(\alpha)}{\partial \beta_{c}}
\end{eqnarray}
where $\beta_{c}=1/k_{B}T_{c}$ is the inverse temperature. On the other hand, the heat exchanges for two isochoric processes BC and DA are given below. The amount of exchanged heats $Q_{BC}$ and  $Q_{DA}$ during isochoric process BC and DA are given by
\begin{eqnarray} \label{heat-B-C}
	Q_{BC} = U_{C} - U_{B}
\end{eqnarray}
and
\begin{eqnarray} \label{heat-C-D}
	Q_{DA} = U_{A} - U_{D} \ .
\end{eqnarray}
By using the partition functions and heat exchanges we can compute work $W$ and efficiency $\eta$.


\section{Work and efficiency of the fractional heat engine}

The work for the Striling like cycle can be obtained from
\begin{eqnarray} \label{work-F}
	W(\alpha) = k_{B} T_{h} \ln \frac{Z_{B}(\alpha)}{Z_{A}(\alpha)} - k_{B} T_{c} \ln \frac{Z_{C}(\alpha)}{Z_{D}(\alpha)} 
\end{eqnarray}
and the efficiency is given by
\begin{eqnarray} \label{efficiency}
	\eta (\alpha) =  1 + \frac{ Q_{BC} + Q_{CD} }{ Q_{DA} + Q_{AB} }
\end{eqnarray}
which is defined in terms of the heat exchanges. In the calculation, for simplicity we neglect some extra energy cost for instance needed energy for insertion and removal of the barrier and needed the energy for the coupling or decoupling of the system to the heat baths. 
\begin{figure} [h!]
	\centering	
	\includegraphics[width=8cm,height=7.5cm]{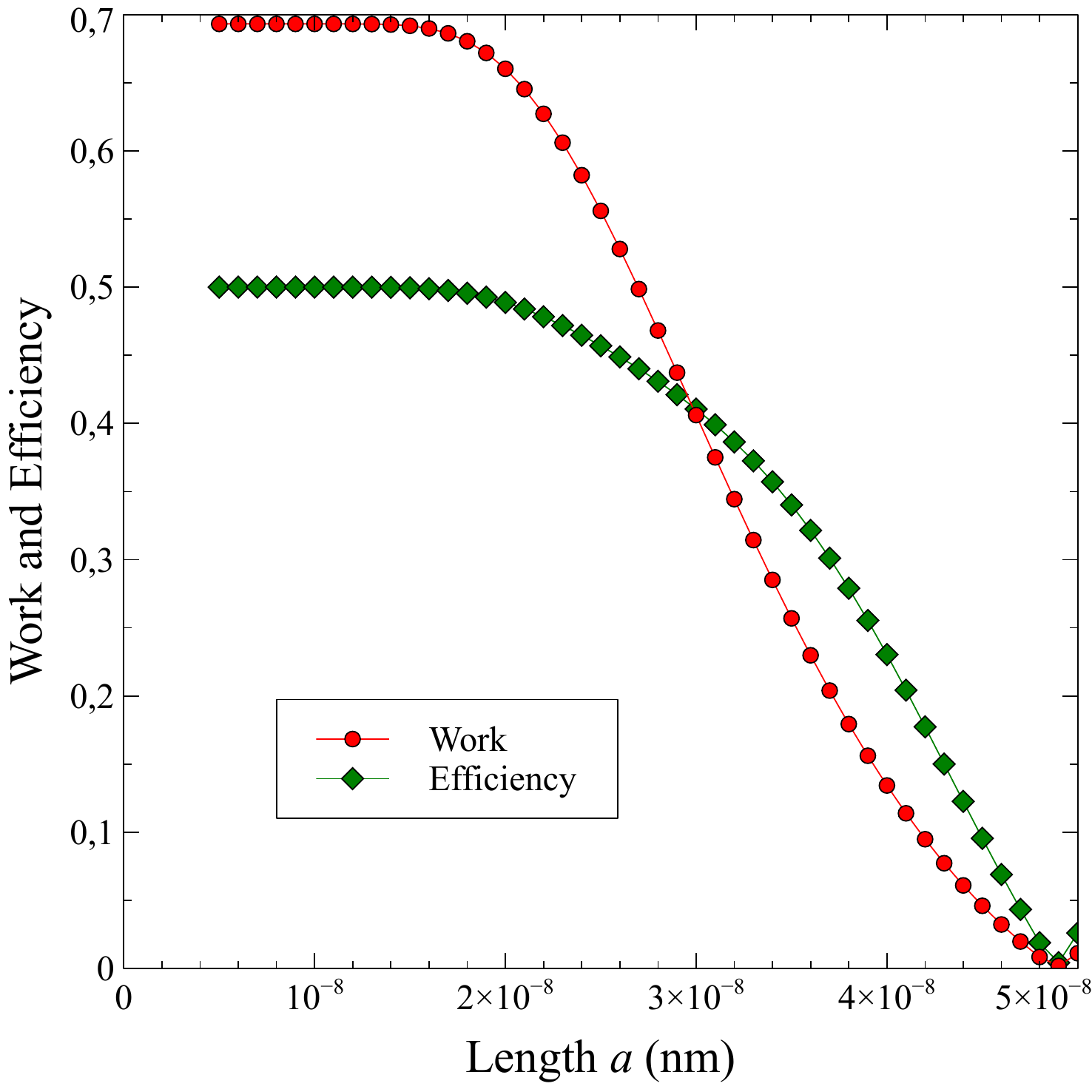}	
	\caption{Work and efficiency of the Szilard quantum heat engine for the non-fractional infinite well.  Here, we have set the constants as $m=9.11\times10^{-31}$ kg, $k_{B}=1.380649\times 10^{-23}$ J/K, $h=6.62607015\times 10^{-34}$ Js $T_{h}=2$ K and $T_{c}=1$ K.}
	\label{h-engine}
\end{figure}

As a result, by using Eq.\,(\ref{work-F}) and  Eq.\,(\ref{efficiency}) we numerically obtained the work $W$ and the efficiency $\eta$ of a Stirling-like cycle presented in Fig.\,\ref{f-cycle} for the $\alpha=2.0$. We note that for the $\alpha=2$, Eq.\,(\ref{f-energy}) reduce to
\begin{eqnarray} \label{n-energy}
	E_{n} =      \frac{n^{2} \pi^{2} \hbar^{2}} {8m a^{2}} 
\end{eqnarray}
which is the energy eigenvalue of a particle in the infinite quantum well. The obtained numerical results for the $\alpha=2.0$ are given in Fig.\ref{h-engine}. Both thermodynamic quantities in Fig.\ref{h-engine} are plotted as a function of the parameter $a$ which denotes the width of the  the potential well. Here, we have set the constants as $m=9.11\times10^{-31}$ kg, $k_{B}=1.380649\times 10^{-23}$ J/K, $h=6.62607015\times 10^{-34}$ Js, $T_{h}=2$ K and $T_{c}=1$ K. 
\begin{figure} [ht!]
	\centering	
	\includegraphics[width=8cm,height=7.5cm]{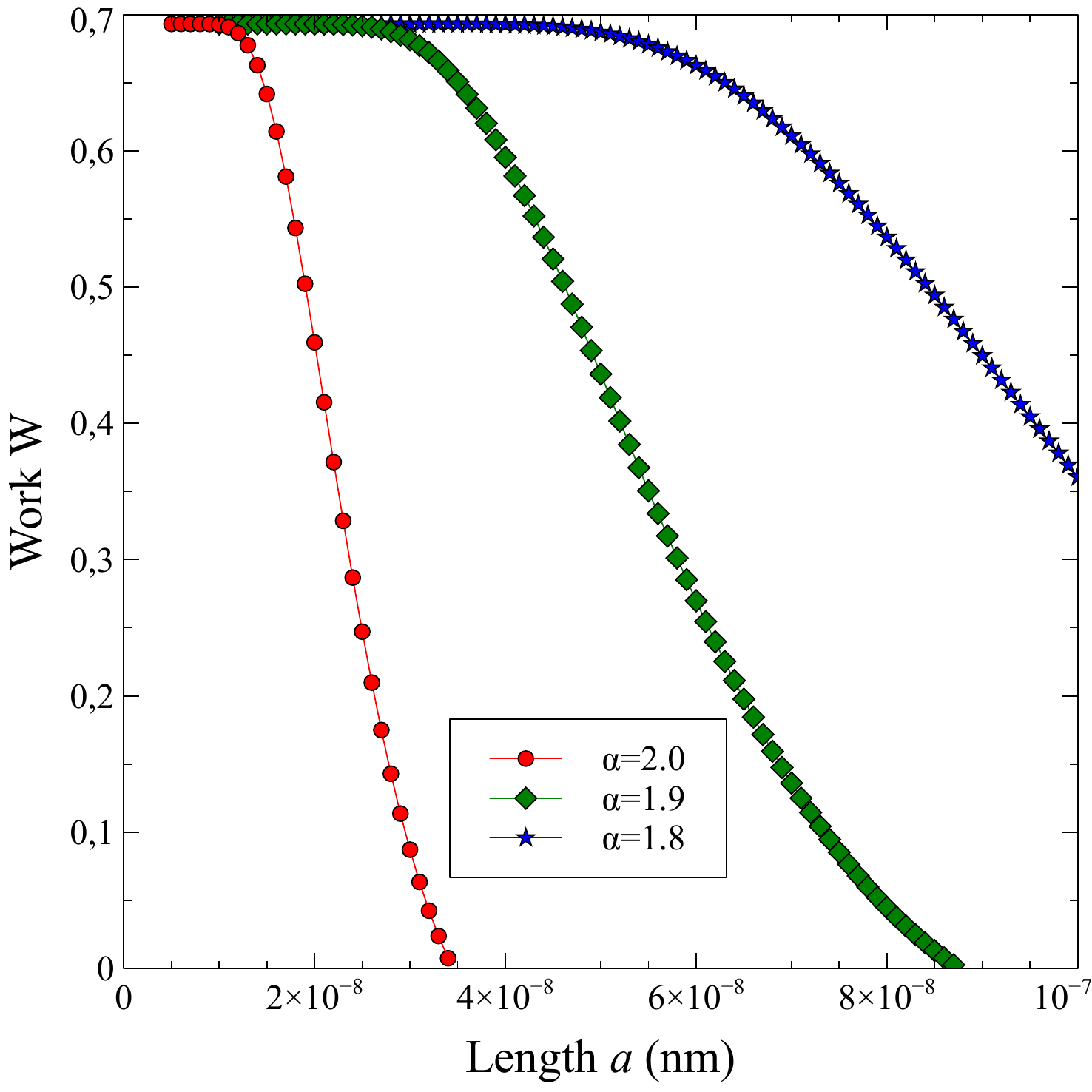}	
	\caption{Plot of the work $W$ of the Szilard quantum heat engine versus length $a$ (in nano-meters) for the various fractional exponents. Here, we have set the constants as $m=9.11\times10^{-31}$ kg, $k_{B}=1.380649\times 10^{-23}$ J/K, $h=6.62607015\times 10^{-34}$ Js, $T_{h}=2$ K and $T_{c}=1$ K. }
	\label{f-work}
\end{figure}
\begin{figure} [ht!]
	\centering	
	\includegraphics[width=8cm,height=7.5cm]{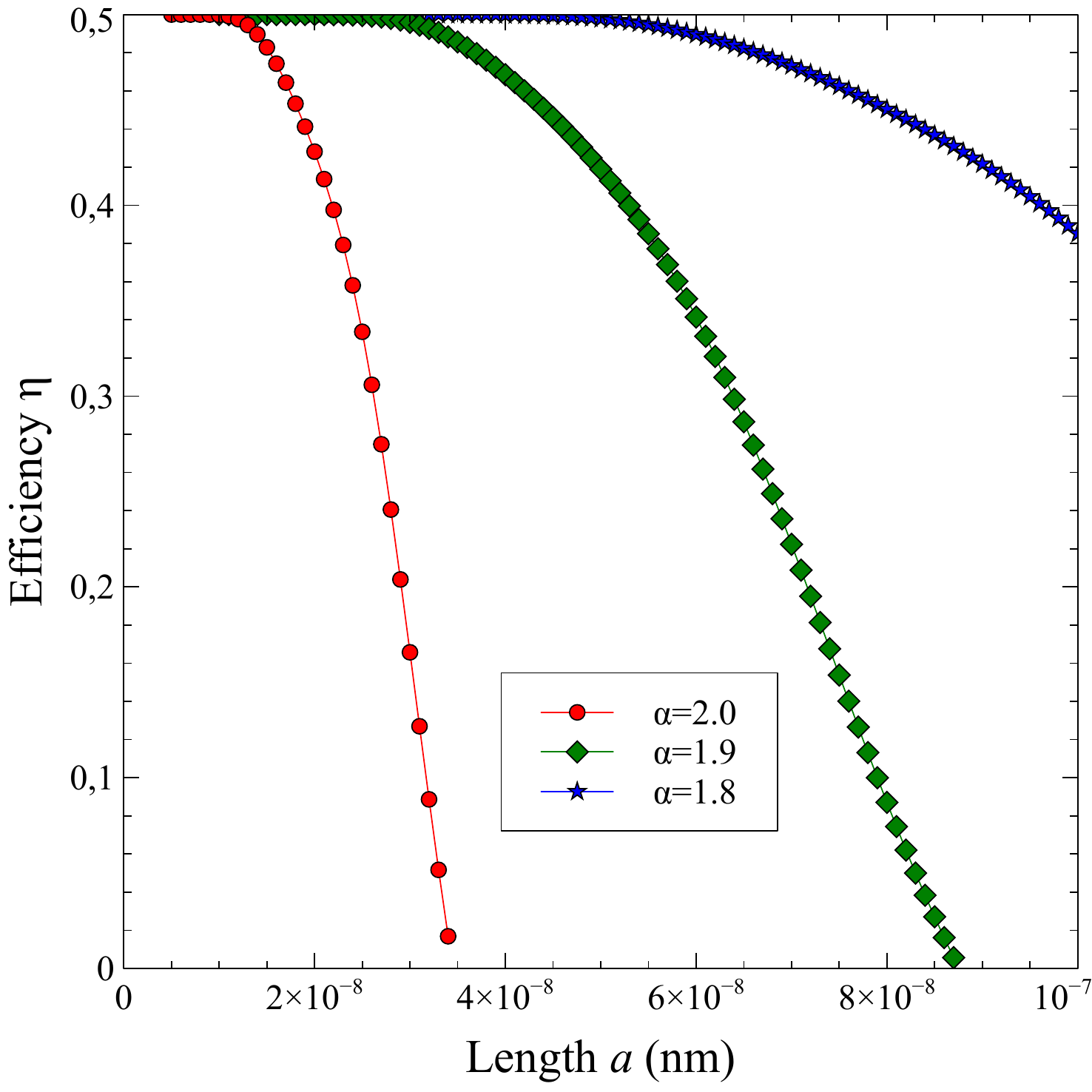}	
	\caption{Plot of the efficiency $\eta$ of the Szilard quantum heat engine versus length $a$ (in nano-meters) for the various fractional exponents. Here, we have set the constants as $m=9.11\times10^{-31}$ kg, $k_{B}=1.380649\times 10^{-23}$ J/K, $h=6.62607015\times 10^{-34}$ Js $T_{h}=2$ K and $T_{c}=1$ K. }
	\label{f-efficiency}
\end{figure}
For $\alpha=2.0$ which denotes the non-fractional infinite well, the obtained results in Fig.\,\ref{h-engine} is compatible with the results in Ref.\,\cite{Thomas2019}. However, one can see that these solutions clearly deviates in the case of the fractional well as seen from Figs.\,\ref{f-work} and \ref{f-efficiency}.

To discuss the effect of the fractional parameter on the work and efficiency we examine the solutions for the $\alpha<2.0$. 

In Fig.\,\ref{f-work} we presented the work curves versus the well width parameter $a$ for various fractional exponent $\alpha$. One can see from the figure that work $W$ is positive and maximum value of work is compatible te result of Fig.\,\ref{h-engine} and the result in Ref.\,\cite{Thomas2019}. When the $a$ increases the extracted work smoothly decreases. However, interestingly, we see that when the space-fractional exponent $\alpha$ decreases, the work persistently lives over longer distances $a$. 

Similarly, we presented the efficiency curves versus the well length parameter $a$ for various fractional exponent $\alpha$ in In Fig.\,\ref{f-efficiency}. One can see a similar behaviour in the efficiency curves. In fact, the efficiency is positive for large distance values for lower values of the fractional exponent $\alpha$. 
Both results in Figs.\,\ref{f-work} and \ref{f-efficiency} indicate that space-fractional quantum engine behaves ve different from conventional quantum Szilard engine.

As a result, in this study, we obtain work and efficiency depending on the potential width of the space-fractional quantum Szilard engine with single-particle for infinite well. We show that the fractional exponent plays important role on the work and efficiency. Additionally, we see that work and efficiency live for large value $a$ when fractional exponent $\alpha$ values become smaller which indicates an increase disorder in the system in the fractional quantum Szilard engine

\section{Conclusion}

The classical and quantum version of the Szilard engine has been studied in detail theoretically and experimentally  \cite{Kim2011,Szilard1929,Kim2012,Li2012,Cai2012,Zhuang2014,Bengtsson2018,Park2013,Thomas2019,Koski2014,Serreli2007,Raizen2009,Bannerman2009,Koski2015,Koski2014b,Koski2014b,Rosnagel2016,Tercas2017,Jarillo2016,Long2016,Erman2016,Martinez2016,Dinis2016,Agarwal2013,Aydiner2021}. These studies show that it is possible to set quantum Szilard heat engines which can extract positive work in nano-size without violate the second law of thermodynamics. On the other hand, these studies prove the connection between thermodynamic entropy and information entropy. 

Nowadays, the interest in these engines is increasing due to their potential technological applications. Obtaining higher work and efficiency in such engines are an important research topic. Therefore, alternative searches have gained importance. In this work, we show that space-fractional quantum heat engines can produce higher efficiency and more robust than conventional quantum heat engines. To prove this, we examined the quantum Szilard heat engine as an example and we presented here the obtained interesting results. In this study, we have not only shown that the high efficiency can be achieved, but also introduced the concept of the fractional quantum heat engine to the literature.



\bibliography{sample.bib}

\end{document}